\documentstyle[aps,draft,epsf]{revtex} 

\def\lesim{\,{\raise-3pt\hbox{$\sim$}}\!\!\!\!\!{\raise2pt\hbox{$<$}}\,}

\def\g{\gamma}

\def\e{\epsilon}

\def\t{\theta}
\def\vt{\vartheta}
\def\k{k}

\def\j{j}
\def\m{\mu}

\def\p{\psi}

\def\ps{\not \! p}

\def\Jsb{\not \! \bar{J}}
\def\Us{\not \! \Upsilon}
\def\pp{{\mbox{\raisebox{-.05cm}{$\stackrel{\perp}{p}$}}}}
\def\ppm{{\mbox{\raisebox{-.05cm}{$\stackrel{\perp}{p}$}}}
         {\mbox{\raisebox{.01cm}{$^{\;\mu}$}}}}

\def\ppdo{{\mbox{\raisebox{-.05cm}{$\stackrel{\perp}{p}$}}}
         {\mbox{\raisebox{.01cm}{$^{\; 2}$}}}}

\def\pps{\not \! {\mbox{\raisebox{-.05cm}{$\stackrel{\perp}{p}$}}}}
\def\ls{\not \! \lambda}
\def\ov{\over}

\title{Majorana Neutrinos and Gravitational Oscillation.}
 \vspace{1.3cm}
\author{Mou Roy and Jos\'{e} Wudka\\ }
\address{Department of Physics \\
         University of California, Riverside \\
        California 92521-0413, U.\ S.\ A.  \\ }
\preprint{UCRHEP-T174}
\date{\today}
\begin{document}

\maketitle

\begin{abstract}
We analyze the possibility of
encountering resonant transitions of high energy Majorana neutrinos
produced in Active Galactic Nuclei (AGN). We consider gravitational,
electromagnetic and matter effects and show that the latter are
ignorable. Resonant oscillations due to the gravitational interactions
are shown to occur at energies in the PeV range for  magnetic
moments in the $ 10^{-17} \mu_B $ range. 
Coherent precession will dominate for larger magnetic moments.
The alllowed regions for gravitational resonant transitions
are  obtained.

\end{abstract}
 
%\newpage

\bigskip\bigskip

\section{Introduction}

Majorana particles are natural representations of massive neutrinos
since the  most general mass term for a four component fermion field
describes two Majorana particles with different masses. Majorana
neutrinos also appear in many extensions of the minimal Standard Model;
this is the case, for example, in  SO(10) grand unified
theories~\cite{so}.

Neutrinos in general, and in particular Majorana neutrinos, can be used to
probe the core of some of the most interesting cosmological objects. Due
to their small cross sections these particles can stream out unaffected
from even the most violent environments such as those present in Active
Galactic Nuclei (AGN). The presence of several neutrino flavors and spin
states modifies this picture: in their trek from their source to the
detector the neutrinos can undergo flavor and/or spin transitions which
can obscure some of the features of the source. Because of this, and due
to the recent interest in neutrino astronomy (e.g. AMANDA, 
Nestor, Baikal, etc.~\cite{jw}), it becomes important to
understand the manner in which these flavor-spin transitions can occur,
in the hope of disentangling these effects from the ones produced by the
properties of the source. Without such an understanding it will be
impossible to determine the properties of the AGN core using solely the
neutrino flux received on Earth.

In a previous publication~\cite{mou} we considered the effects of the
AGN environment on the neutrino flux under the assumption that all
neutrinos were of the Dirac type. In this complementary publication we
will consider the case of Majorana neutrinos and provide a deeper
phenomenological study of this system, concentrating on the dependence
of the effects on the magnitude of the neutrino magnetic moment and on
the energy dependence of the predicted neutrino fluxes.

The evolution of Majorana neutrinos in  AGN is influenced both by its
gravitational and electromagnetic interactions. The latter are
due to the coupling with the  magnetic field through  a transition magnetic
moment ~\footnote {Majorana neutrinos are self conjugate particles which
implies 
that the flavor diagonal magnetic moment form factor vanishes identically
~\cite{kayser}. 
The absence of flavor-diagonal magnetic moments 
reduces the number of possible flavor and/or spin resonances that
can be present compared to the Dirac case.}. The combination of these
effects leads to  $ \nu_\mu \rightarrow \bar{ \nu_e } $ or $\bar{ \nu_\tau}$
transitions. For simplicity we will deal with two neutrino species only,
the extensions to three (or more) species is straightforward (though the
analysis can become considerably more complicated~\cite{folgi}).

The paper is organized as follows. We start with a brief description
of neutrino production and gravitational oscillation in AGN environment
in section 2. This is followed by a calculation of transition and survival
probabilities of oscillating  neutrinos and the resulting flux
modifications (sections 3 and 4). In section 5 we give our conclusions.

\section{Neutrino Oscillations in AGN }

Active Galactic Nuclei
(AGN) are the most luminous  objects in the Universe, their 
luminosities ranging from $10^{42}$ to $10^{48}$ ergs$/$sec. 
They are believed to be powered by a central supermassive black hole
whose masses are of the order of $10^4 $ to $10^{10} M_{\odot}$. 

High energy neutrino
production in an AGN environment can be described using the
so-called  spherical accretion
model~\cite{sph} from which we can estimate the matter density
and also the magnetic field (both of which are needed to study the
evolution of the neutrino system). Neutrino production in this model
occurs via the $ \pi ^\pm \rightarrow \mu^\pm
\rightarrow e^\pm$ decay chain~\cite{sph,zp} with the pions being
produced through the collision of fast protons (accelerated through
first-order diffusive Fermi mechanism at the shock~\cite{sph}) and the
photons of the dense ambient radiation field.
These neutrinos are expected to dominate the neutrino sky
at energies of 1 TeV and beyond~\cite{zp}.

Within this model the order of magnitude of the
matter density $\rho$ for typical cases
can be estimated at $10^1-10^4 \;{\rm eV}^4$. The magnetic field, for 
reasonable parameters, is of the order of $ 10^4 $ G ~\cite{mou,zp,bbr}.

In order to determine the effective interactions of the Majorana
neutrinos in an AGN environment we start, following~\cite{mou}, from
the Dirac equation in curved space including their weak and
electromagnetic interactions ~\footnote{ The vector current is zero
for Majorana neutrinos; they interact with matter only through the
axial vector current unlike Dirac neutrinos.}
\begin{equation}     
[i e^\mu_a \gamma^a (\partial_\mu + \omega_\mu) - m + \not\! J \gamma_5 
+ \mu \sigma^{ab} F_{ab}] \p = 0 , 
\label{aaaa} 
\end{equation}
where $e^\mu_a$ are the tetrads, $ \gamma^a $ the usual Gamma matrices,
$m$ is the mass matrix, $ \not \! J = J_a \gamma^a $ denotes the
weak interaction current matrix, $\mu$ is the neutrino magnetic moment matrix,
$F^{ab}$ the electromagnetic field tensor, $ \sigma^{ab} = { 1 \ov 4} [ \g_a,
\g_b ] $;
 and the spin connection equals 
\begin{equation}
\omega_\mu={1\over 8}[\gamma_a,\gamma_b] e^{\nu a} e^b_{\nu;\mu}   \label{spinco}
\end{equation}
where the semicolon denotes a covariant derivative. 
We used Greek indices ($\mu, \, \nu , \ldots $) to denote space-time 
directions, and Latin indices ($a , \, b , \ldots $) to denote
directions in a local Lorentzian frame.

The method of extracting the effective neutrino Hamiltonian from
(\ref{aaaa}) is studied in detail in~\cite{mou}. We will therefore
provide only a brief description of the procedure for completeness.

The first step of the semiclassical approximation is to consider a
classical geodesic $ \bar x^\mu (l)$ parameterized by an affine
parameter $l$. Along this curve we construct three vector fields $
\nu^\mu_A (l) , \, A=1,2,3$ such that $ \bar x + \nu_A \xi^A $
satisfies the geodesic differential equation to first order in the $
\xi^A $. We then use $ \{ l , \xi^A \} $ as our coordinates
(see, for example,~\cite{st}).

Next we consider the classical action as a function of the coordinates
which satisfies the relation $ p_\mu = \partial S / \partial x^\mu $ ($p$
is the classical momentum), and define a spinor $ \chi $ via the usual
semiclassical relation
\begin{equation}
\psi = e^{i S} \chi \label{defofchi} 
\end{equation}

In our calculations it proves convenient to define a time-like
vector $\ppm$ corresponding to the component of
momentum $p$ orthogonal to the $ \nu^\mu_A$
\begin{equation} \ppm = p^\mu - c_A (N^{-1})^{AB} \nu^\mu_B; 
\qquad N_{AB}= \nu^\mu_A \nu^\nu_B g_{\mu\nu} \end{equation}
It can be shown that $ c_A $ are constants~\cite{mou}.
We 
denote by $ R $  the length-scale of the metric,
so that, for example, $ \omega_\m \sim { 1/R}$; and let $P$ be the
order of magnitude of the momentum of the neutrinos.
We then make a double expansion of $\chi$, first in powers of
$\xi$ and then in powers of $1/pR$. We substitute these expressions 
into the Dirac equation and demand that each term vanish separately.

It is then possible  to reduce the resulting equations to a
Schr\"{o}dinger-like equation 
involving only $\chi$ which reads
\begin{eqnarray} 
i \dot\chi &=& \tilde{H}_{\rm eff}\; \chi  \nonumber \\ 
\tilde{H}_{\rm eff} &=&  -{ 1 \over 2} (\ps + m) \left(
i \gamma^a \bar e^\mu_a \bar \omega_\mu + \Jsb \gamma_5
\right) + { 1 \over 2} m^2 - { i \over {2 \pps}}
\dot{\ps} + {i  \over 2}{\not \! \bar{\Upsilon}}^A {\not \! \bar{\lambda}}_A 
+ \mu_t \sigma^{ab} F_{ab}, \nonumber\\
\label{schr}    
\end{eqnarray}
where, 
\begin{eqnarray}
\ls_A &=& { 1 \over 2} \dot{N}_{AB}\Us^B - \bar{e}^{\mu}_{a;\nu}
p_{\mu} \nu^{\nu}_A\gamma^a, \nonumber \\
\Us^A &=& (N^{-1})^{AB} \left(\nu_{\mu B} - { c_B \pp_{\mu} \over {\ppdo}}\right)
\bar{e}^{\mu}_a \gamma^a. \label{poteo} \nonumber
\end{eqnarray}
(the over-bar indicates that the quantities are evaluated on the geodesic:
$ x = \bar x(l) $).
Eqn. (\ref{schr}) determines the evolution of the 8-component (for two flavors) spinor $ \chi $. We are interested, however only in the 4-components
representing the spinors with positive momenta~\cite{mou}
({\it i.e.} those directed from the source to the observer). Projecting
$ \tilde{H}_{\rm eff} $ into this 4-dimensional subspace
yields the effective Hamiltonian $ H_{\rm eff}$
for the states of interest
\begin{equation} 
\ H_{\rm eff} = i \dot{\alpha} + { 1 \over 2} m^2 + 
p \cdot J_{\rm eff} \, 
\left(\begin{array}{cc}
1 & 0 \\ 0 & -1
\end{array} \right)
+ \mu_t \sqrt{ \pp \cdot p } \, 
\left(\begin{array}{cc}
0 & B^* \\ B & 0
\end{array} \right)
\label{heff}
\end{equation}
where $B$ denotes the magnetic field and
the effective current $J_{\rm eff}$ is defined by
\begin{equation} 
J_{\rm eff}^a = J_W^a - J_G^a \label{jefa}   
\end{equation}
in terms of the weak-interaction current $J_W^a$ and  the ``gravitational current"
\begin{equation}
J_G^a = { 1 \over 4 } \e^{abcd} \bar{\lambda}_{fcd} 
\left(\eta^f_b + { 2 p^f \pp_b \over \ppdo} \right)  ,    \label{jefb}  
\end{equation}
where
\begin{equation}
\lambda_{fcd} = (e_{f\mu ,\nu} - e_{f\nu,\mu}) e^{\mu}_c e^{\nu}_d,
\qquad \eta^{ab} = diag(1,-1,-1,-1) .\label{you}
\end{equation}
The term  $ \dot{\alpha} $ in (\ref{heff}) is flavor and spin diagonal
and can be eliminated by redefining the overall
phase of $ \chi $ with no observable consequences.

The evolution of
Majorana neutrinos through matter in the presence of strong gravitational
fields
incorporating magnetic effects 
is thus,
\begin{equation}
i {d \over dl} \left(\begin{array}{c}
 \nu_{e}\\ \nu_{\mu}\\ {\bar\nu_{e}}\\ {\bar\nu_{\mu}}
\end{array}
\right) =  H_{\rm eff} \left(\begin{array}{c}
\nu_{e}\\ \nu_{\mu}\\ {\bar\nu_{e}}\\ {\bar\nu_{\mu}}
\end{array}\right)
\end{equation}
where $l$ is the affine parameter ~\footnote{Note that the left hand side
involves differentiation with respect to the affine parameter
which has units of  $ \hbox{(mass)}^{-2}$. 
Therefore $ H_{\rm eff}$ has units of $\hbox{(mass)}^2$ which differs
from the usual Hamiltonian units. In cases where the neutrino energy $E$ is
conserved the usual effective Hamiltonian is    $ H_{\rm eff}/E$.
}
and
$ H_{\rm  eff}$ is  the $4 \times 4$ matrix 
containing the effects of
the weak, electromagnetic and gravitational neutrino interactions,
explicitly
\vspace{0.5cm}
\begin{equation} 
H_{\rm eff}=
\left(\begin{array}{cccc}
      \vspace{0.5cm}
p \cdot J_{\rm eff} &
{1\over 4} \Delta m^2 \sin 2\vt  &
0&
E \mu_t B^\ast \\
      \vspace{0.5cm}
{1\over 4} \Delta m^2 \sin 2\vt  &
\qquad p \cdot J_{\rm eff} + {1\over 2} \Delta m^2 \cos 2\vt &
-E \mu_t B^\ast &
0\\
\vspace{0.5cm}
0&
-E \mu_t B&
-p \cdot J_{\rm eff} &
{1\over 4} \Delta m^2 \sin 2\vt\\
%      \vspace{0.5cm}
E \mu_t B  &
0 &
{1\over 4} \Delta m^2 \sin 2\vt&
\qquad -p \cdot J_{\rm eff} + {1\over 2} \Delta m^2 \cos 2\vt \\
\end{array} \right)
\vspace{0.5cm}
\label{gau}
\end{equation}
where $\vt$ is the neutrino mixing angle, $ \Delta m^2 = m^2_1 - m^2_2 $, 
 $E$ the energy 
of the particle, $B=B_1 + i B_2$ where $ B_{1,2} $ are
the AGN magnetic field components perpendicular to the
direction of motion, and
$\mu_t$ the
transition magnetic moment. 
It is worth noting that, in contrast to the Dirac case, 
antineutrinos exhibit matter interactions. This Hamiltonian includes
gravitational as well as electroweak effects; the latter have been
studied previously (see, for example,~\cite{lm}).

In our calculations we will use the Kerr metric to allow for the
possibility of rotation of the central AGN black hole (we also have assumed
that the accreting matter generates a small perturbation of the 
gravitational field). The metric for a Kerr black hole contains two
parameters, $ r_g $, the horizon radius
and $a$  the total angular momentum of the black hole per unit mass.
The geodesics in this gravitational field have three constants of the
motion, commonly denoted by $E$, $L$ and $K$. The first corresponds to
the energy, the second to the angular momentum along the black-hole
rotation axis; the third constant $K$ has no direct interpretation, but
is associated with the total angular momentum~\cite{kerrref}.

Using the AGN models mentioned above we can compare the weak interaction
current $J_W$, to $J_G$, its gravitational counterpart. 
The orders of magnitude are,
\begin{equation}
J_W  =  {G_F \; \rho \over m_p} \sim \left(10^{-33} \; \rho_{\rm eV} \right) \;\; {\rm eV}^{-1}
,\qquad J_G \sim R^{-1}
\end{equation}
where  $ G_F $ is the Fermi coupling constant, $m_p$ the proton rest mass and
$\rho_{\rm eV}$ is the density in $ {\rm eV}^4$ units which for 
typical cases is $10^1-10^4 \;$~\cite{zp}. Taking $ R \sim r_g $, 
the  gravitational current part is found to dominate
the weak current part for all relevant values of $ r_g  \; (10^{14}\; {\rm to}
\; 10^{20} {\rm eV^{-1}})$.
In the following we will drop $ J_W $.

Setting $ J_W = 0 $, $ J_{\rm eff}$ can be written in terms
of a dimensionless function $f$,
\begin{equation}
\vert p\cdot J_{\rm eff} \vert = Er_g^{-1} f(r/r_g,\t,\j,\k, a/r_g).
\label{resca}
\end{equation}
where we have chosen normalized parameters 
\begin{equation} \j ={ L \over E r_g } , \qquad \k={ K \over (E r_g)^2}
\label{theparameters}
\end{equation}
We have plotted  $ \vert p \cdot J_{eff} \vert $
in figures~\ref{fig:f.1} and~\ref{fig:f.2} for some typical parameter values.

Using (\ref{gau}) we can determine the AGN regions where resonant
transitions occur. These
resonances are governed by the $ 2 \times 2$ submatrices of
(\ref{gau}) for each pair of states. The two possible
resonances~\footnote{The two remaining transitions have matrices with
vanishing off-diagonal elements and will not exhibit resonances.}
are obtained by equating the diagonal terms for
each submatrix and give rise to the following resonance conditions

\begin{eqnarray}
{1\over4} \Delta m^2 \cos 2\vartheta &=&  p \cdot J_G\quad \qquad 
(\nu_e\; \rightarrow \;\bar{\nu}_\mu) 
\label{resa} \\ \nonumber
{1\over4} \Delta m^2 \cos 2\vartheta &=& -p \cdot J_G\quad \qquad 
(\nu_\mu \; \rightarrow \; \bar{\nu}_e) 
\label{resb}
 \end{eqnarray}

As can clearly be seen from the above two equations the resonant transitions
do not occur simultaneously.
We have considered electron and muon neutrinos, similar results hold for
any other pair of flavors. 

\setbox4=\vbox to 160 pt {\epsfysize= 5 truein\epsfbox[0 -200 612 592]{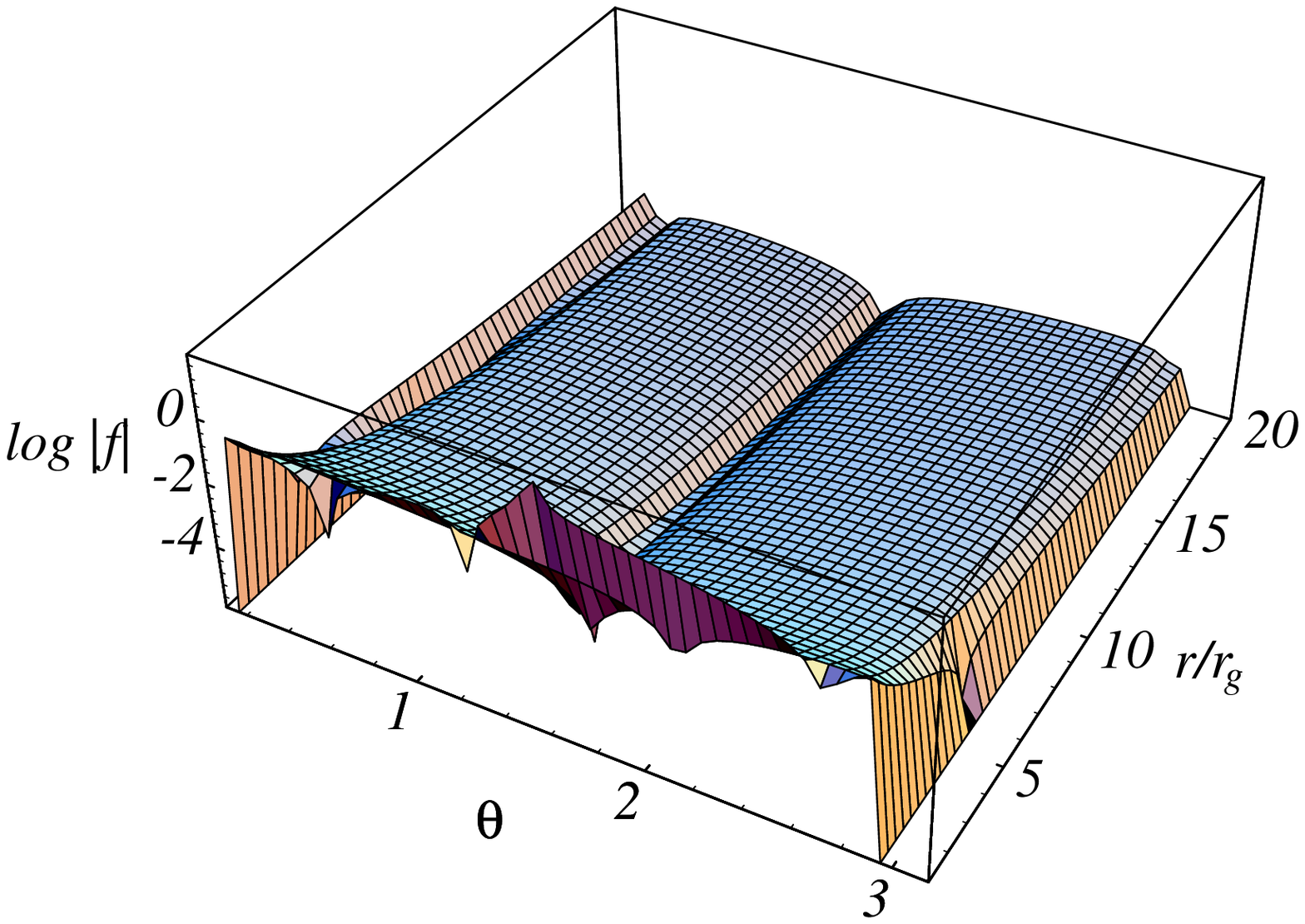}}
\begin{figure}
\centerline{\box4} 
\caption{Plot of $ \ln |f|$, see (\ref{resca}), for $\j=0.15,
\; \k=0.4$, $a/r_g=0.4$ and $M=10^8 M_\odot $. \label{fig:f.1}}
\end{figure}

\setbox4=\vbox to 180 pt {\epsfysize=4 truein\epsfbox[0 -140 612 652]{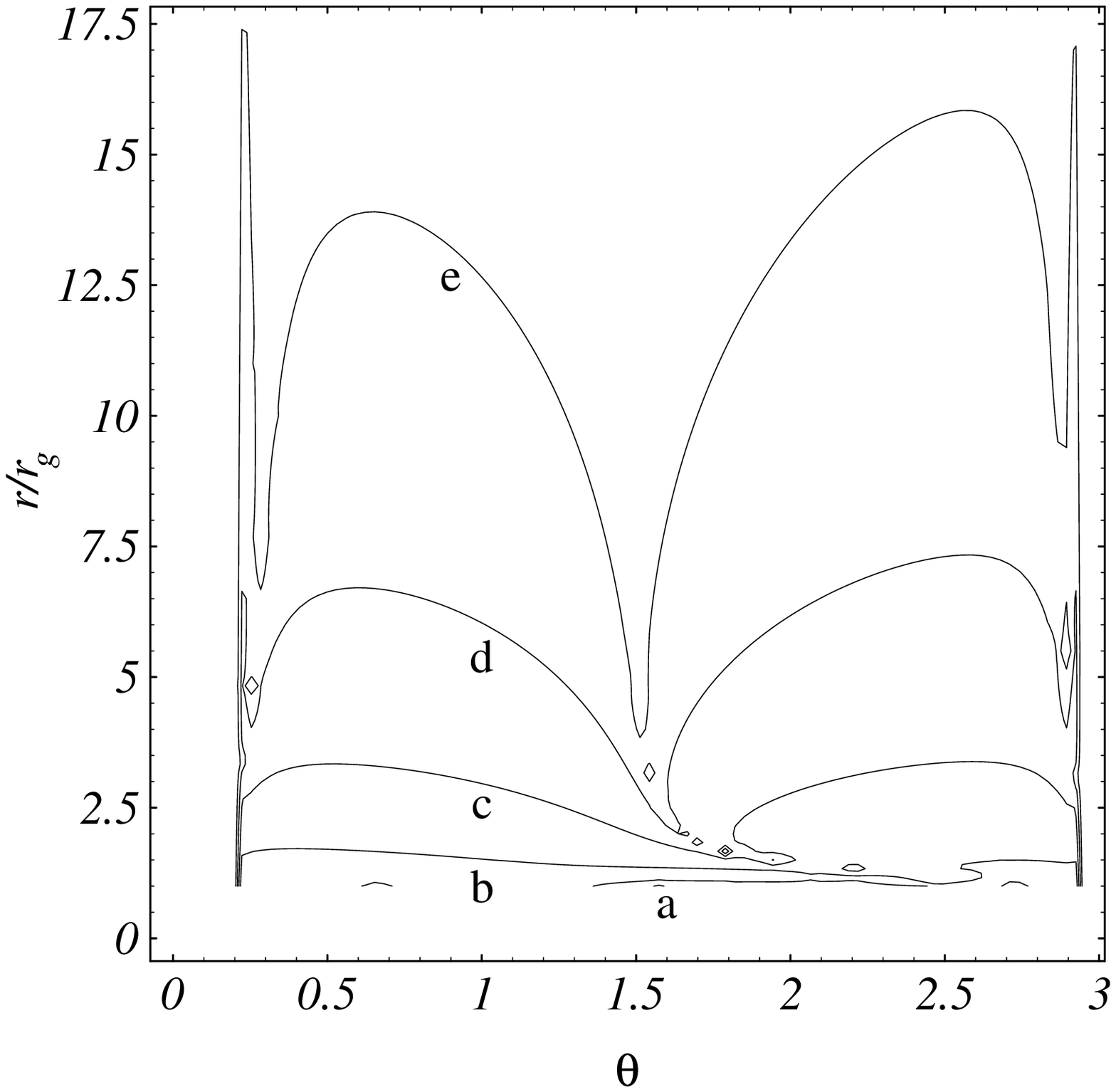}}
\begin{figure}
\centerline{\box4}
\caption{Contour-plot of $f$, see (\ref{resca}), for$\j=0.15,\;
\k=0.4$, $a/r_g=0.4$ for the case $M=10^8 M_\odot $.
The contours a,b,c,d,e correspond respectively to
$f =\pm 1,\, \pm 10^{-1},\, \pm 10^{-2},\, \pm 10^{-3},\, \pm 10^{-4}$,
(positive values for  $0\le \theta \le \pi /2$, negative
values for $\pi /2 \le \theta \le \pi$). \label{fig:f.2}}
\end{figure}

Resonances occur provided $f$ is comparable to $ \pm \Delta m^2 \cos 2\vartheta
 r_g/E $ as can be seen from (\ref{resa}) and (\ref{resb}).
As suggested by figures~\ref{fig:f.1} and~\ref{fig:f.2}, 
we have verified that, just as in the Dirac case~\cite{mou},
for almost all values of $j$, $k$ and $a$ the neutrinos will undergo
resonances at a radius $r$ (the precise value of which changes with $
\theta $) for all relevant values of $ \Delta m^2 $ (in the range $10^{-2}$
to $ 10^{-10}  {\rm eV^2} $) provided $E$ is large enough ($ > 10^8 {\rm eV}$
in this case) . This
implies that practically all Majorana neutrinos will experience
resonances provided their energy is large enough.
As an example for $E \sim 1$ PeV and $\Delta m^2_{12} \sim 10^{-6} $eV$^2$
(Solar large angle solution), the resonance contour corresponds to
(d) in Fig.~\ref{fig:f.2}.

\section{Probabilities for Allowed Transitions}

In this section we evaluate the survival and
transition probabilities of neutrino transitions and look into the
region in parameter space where resonant neutrino transitions occur.
The average probabilities for oscillating 
neutrinos (including non-adiabatic effects)
produced in a region with mixing angle $ \vt_G $, and detected 
in vacuum where the mixing angle is $\vt$,
are given in general by~\cite{{parke},{panta}}

\begin{equation}
P(\nu_\mu \rightarrow \nu_\mu) = 
{1 \ov 2} + \left({1 \ov 2} - P_{LZ} \right)
\cos 2 \vt_G \cos 2 \vt
\label{prob4}
\end{equation}
and
\begin{equation}
P(\nu_\mu \rightarrow \bar{\nu}_e) = {1 \ov 2} - \left({1\ov2} -
P_{LZ} \right) \cos 2 \vt_G \cos 2 \vt
\label{prob2}
\end{equation}
where $ \vt_G $ is the gravitational mixing angle,
\begin{equation}
\tan2\vt_G = \left. { 4 E \, \mu_t B  \ov \Delta m^2 \cos 2\vt + 4 p \cdot J_G
}\right|_{\rm prod.}
\label{theta}
\end{equation}
(evaluated at the production point)
and $\mu_t$ is the transition magnetic moment, $B$ the magnetic field,
$\Delta m^2$ the usual mass difference parameter and $J_G$ given in eqn.(\ref{resca}).
$ P_{LZ} $ is the Landau Zener probability
\begin{equation}
P_{LZ}= \exp \left\{ -2 \pi^2 \, {\beta^2 \over \alpha } \right\}
\end{equation}
where
\begin{equation}
\beta = E \mu_t B \vert_{\rm res}\qquad 
\alpha = { d \ov dl } \left[ - p \cdot J_{\rm eff} 
\right]_{\rm res} \label{alph}
\end{equation}

The condition for  adiabatic resonances to
induce an appreciable transition probability is
$ 2 \pi^2 \beta^2 \ge \alpha $ which in terms of the magnetic moment
implies
\begin{equation}
\mu_t \ge { 1 \over E B} \left| { \Delta
m^2 \over 2 \pi^2 \Lambda } \right|^{1/2} =
\mu^{\rm res}_{\rm min} \label{cond2}
\end{equation}
where $\Lambda = \left| d l / d \ln \left( p \cdot J_{\rm eff} \right)  
\right| $ is a measure of the scale of the gravitational field (divided
by $E$), and we have assumed that the magnetic field
remains constant over an interval of magnitude $ \sim \Lambda $. 
In order to exhibit the energy dependence of $
\mu^{\rm res}_{\rm min} $ we estimate $ \Lambda \sim r_g \,/ E $ whence
\begin{equation}
\mu^{\rm res}_{\rm min}  \sim { 1 \over B} \left| { \Delta
m^2 \over 2 \pi^2 r_g E } \right|^{1/2} 
\end{equation}

Using equations (\ref{prob4}) and (\ref{prob2}) 
we can then calculate the survival and transition
probabilities for the neutrinos. The results depend on the particular
geodesic followed by the neutrinos, that is, it depends on the
parameters $E$ as well as $j$ and $k$ defined in (\ref{theparameters}).
The results are also dependent on the characteristics of the
gravitational field and its environment through the magnetic field, the
horizon radius $ r_g $ and the angular momentum parameter $a$. Finally
there is also an important dependence on the neutrino parameters $
\Delta m^2 $ and $ \mu_t $.

For the study of the probabilities there are two regions of interest
(assuming (\ref{cond2}) is satisfied). If  $ p
\cdot J_G > ( E \: \mu_t B ) $ the system experiences adiabatic
resonances whenever the energy
is such that $ \vt_G \sim \pi/2 $. If $ p
\cdot J_G < ( E \: \mu_t B )  $ the system will exhibit coherent
precession ~\cite{okun} if $  E \: \mu_t B > \Delta m^2 $ or no appreciable
transitions if $  E \: \mu_t B < \Delta m^2 $. 

For a study  of the resonance scenario we will
choose $ \mu_t = 10^{-17} \mu_B $ which is amply allowed by the current
experimental and astrophysical constraints and still ensures neutrino 
adiabatic resonant transitions for all $ \Delta m^2 $ values considered
provided $E$ is sufficiently large~\footnote{
For $ E=1 $PeV and $ \Delta m^2 = 10^{-8} $eV$^2$, $ \Lambda \sim r_g/E $,
we have $ \mu^{\rm res}_{\rm min} \sim 10^{-17} \mu_B $.}. This value is
smaller than the one usually considered in the literature due to our
being concerned with energies in the PeV range and the fact that 
$ \mu^{\rm res}_{\rm min} $ decreases as $ 1/ \sqrt{E} $.
For larger values of $ \mu_t $ transitions are dominated by coherent 
precession (provided $ E \mu_t B > \Delta m^2 $).

Using  expressions (\ref{prob4}) and (\ref{prob2})
we obtained the
probabilities of gravitationally induced adiabatic transitions, the
results are presented in Fig.~\ref{fig:f.3}.
If $E = E_R$ is the resonant energy (for a particular
choice of parameters) then at that energy $\cos2\vt_G \sim 0$. If $E < E_R$
the transition probability
approaches the vacuum value $  P( \nu_\mu \rightarrow \bar \nu_e ) = 
\sin^2 2\vt $. For $E > E_R$, and for the value of $ \mu_t
$ chosen, $\vt_G \rightarrow \pi/2$
and the transition probability approaches 
$ P( \nu_\mu \rightarrow \bar \nu_e ) = \cos^2\vt$, 
which is $ \sim 1 $; this is our region of interest.
Equation (\ref{cond2}) is satisfied in Fig.~\ref{fig:f.3}
when $ P( \nu_\mu \rightarrow \bar \nu_e )$ approaches 1 and $ P( \nu_\mu \rightarrow \bar \nu_\mu)$ approaches zero which corresponds to conditions for
adiabatic resonant transitions.
The energy $ E_R$ denotes the adiabatic threshold energy between no
gravitational effects and adiabatic resonant conversion. This behavior
is illustrated in Fig.~\ref{fig:f.3}. 
Fig.~\ref{fig:f.4} shows the dependence of
the probabilities on $r$ and $ \theta $.

\setbox4=\vbox to 180 pt{\epsfysize=5.5 truein\epsfbox[-100 -220 512 572]{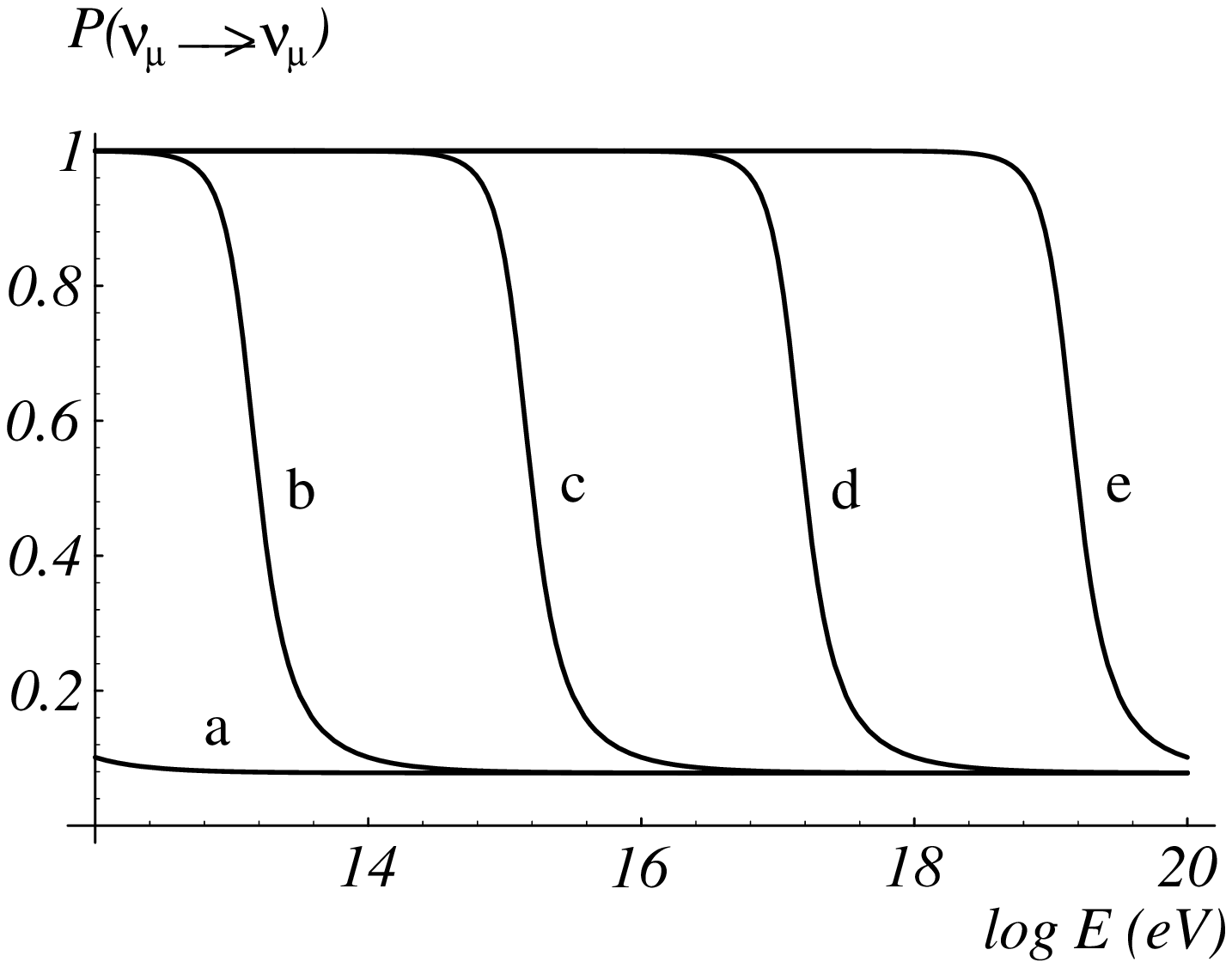}}
\setbox5=\vbox to 180 pt{\epsfysize=5.5 truein\epsfbox[20 -220 632 572]{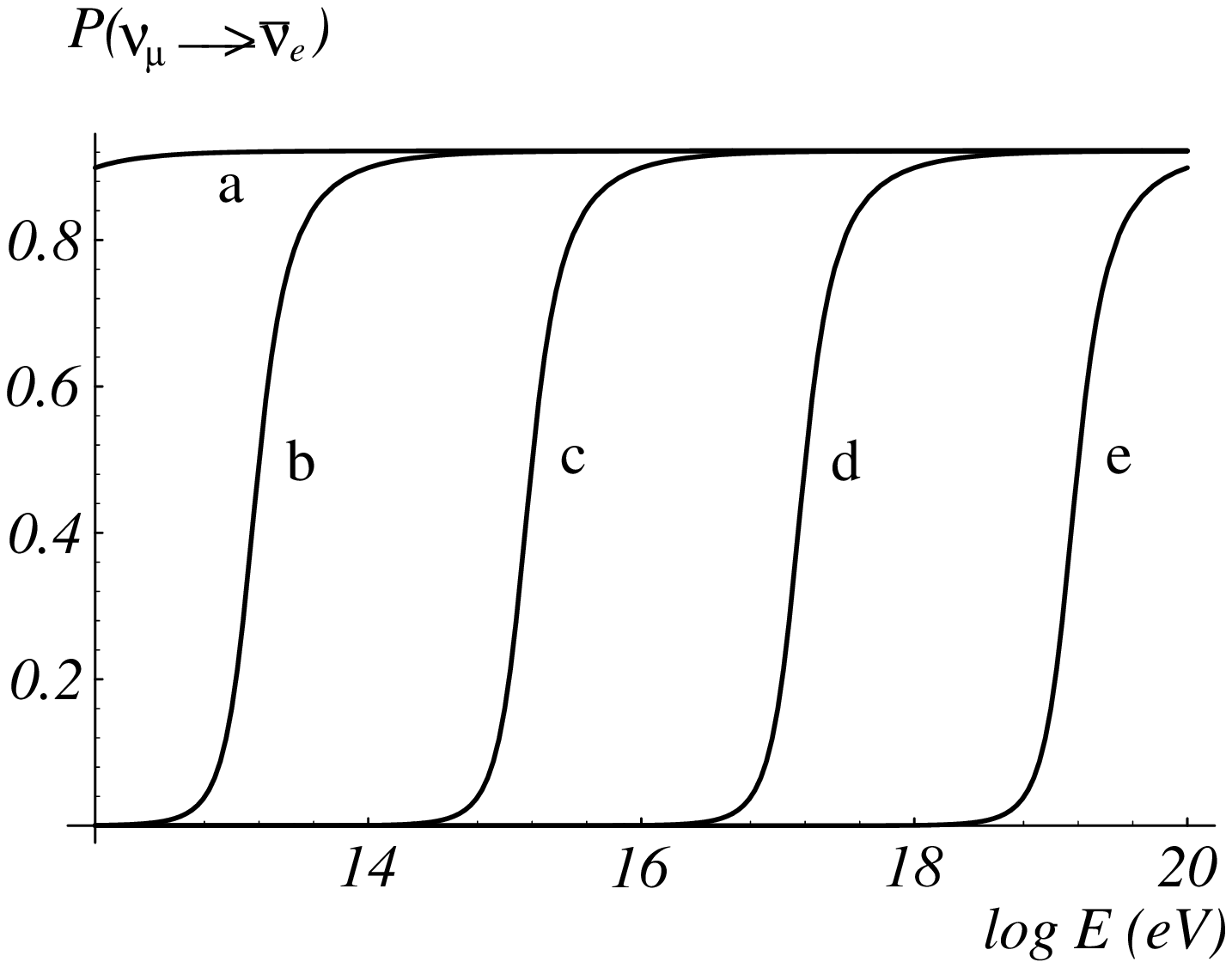}}
\begin{figure}
\centerline{\hfill \box4 \box5 \hfill }
\caption{Plot of the persistence (left) and transition (right)
probabilities as functions
of $E (\rm eV)$ for $\mu_t = 10^{-17} \mu_B $, $r/r_g = 6$, 
$\theta=\pi/2$, $ \j=0.15 $, $\k=0.4$, 
$ a/r_g=0.4$ and $ M = 10^8 M_\odot$.
The curves a,b,c,d,e correspond to $\Delta m^2 = 10^{-10}, 10^{-8}, 10^{-6},
10^{-4}, 10^{-2} {\rm eV}^2 $ respectively. 
\label{fig:f.3}}
\end{figure}

\setbox4=\vbox to 170 pt{\epsfysize=4.5 truein\epsfbox[-30 -230 582 562]{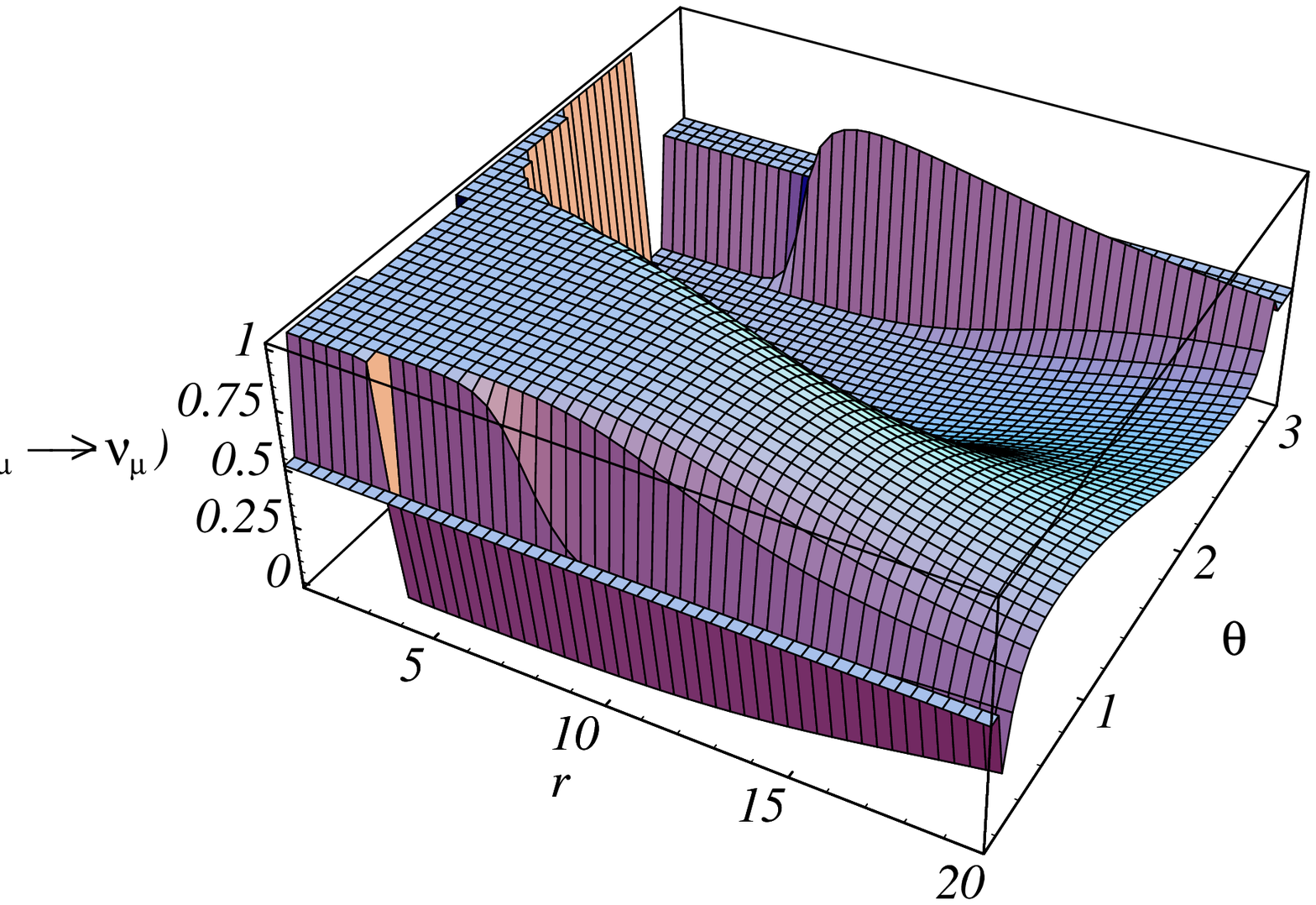}}
\setbox5=\vbox to 170 pt{\epsfysize=4.5 truein\epsfbox[150 -230 762 562]{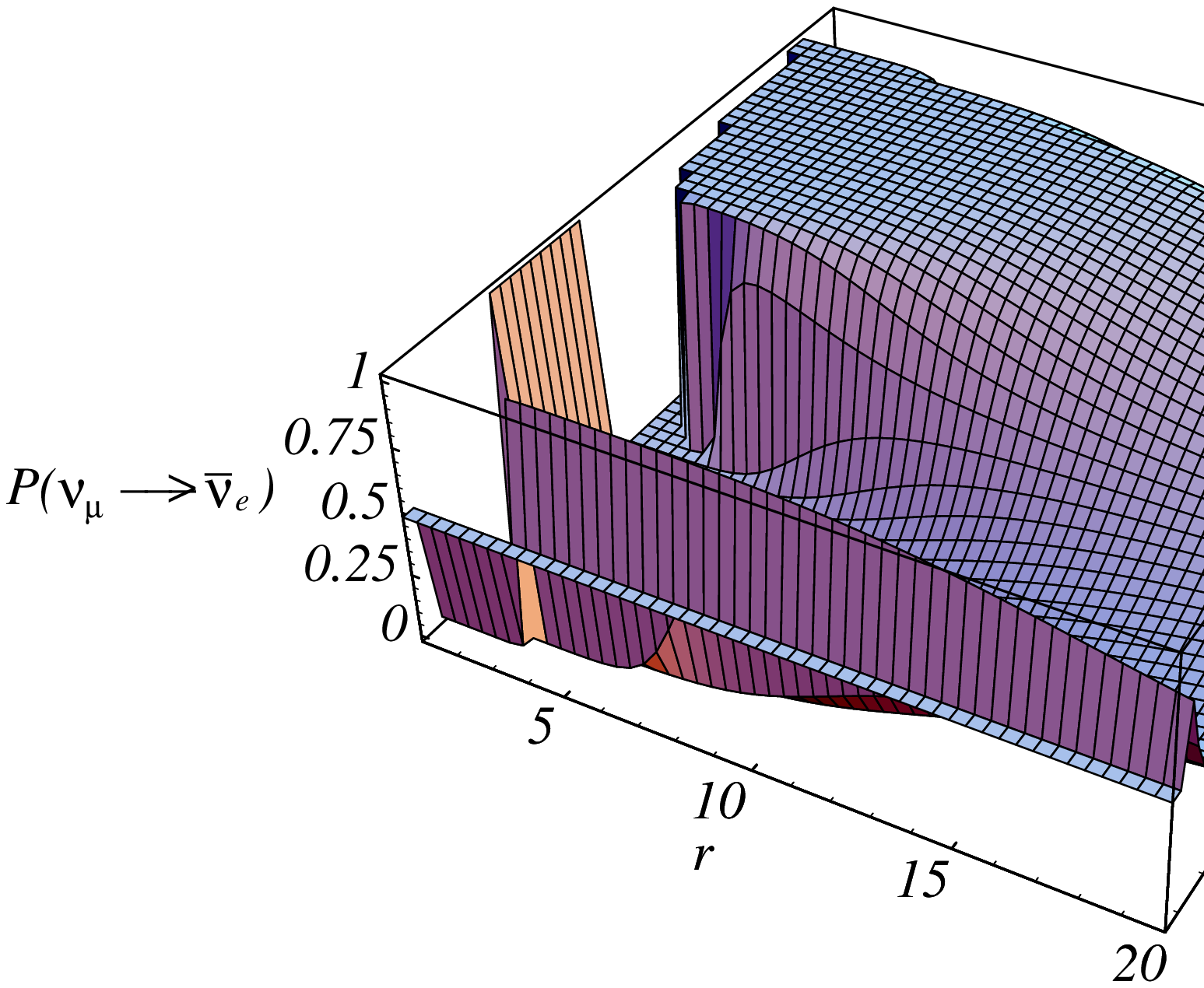}}
\begin{figure}
\centerline{\hfill \box4 \hfill \box5 \hfill}
\caption{Plot of the persistence (left) and transition (right) probabilities 
as functions of
$(r,\theta)$ for $\mu_t = 10^{-17} \mu_B $, $ \j=0.15 $, $\k=0.4$, $a/r_g=0.4$,
$E = 1 {\rm PeV}$, $\Delta m^2 = 10^{-8}{\rm eV}^2 $ and $ M = 10^8 M_\odot$.
 \label{fig:f.4}}
\end{figure}

The region in parameter space where adiabatic resonances occur is
restricted by $ \mu_t > \mu^{\rm res}_{\rm min} $, $\left|  p \cdot J_G 
\right| > E
\mu_t B $ and by $ E > E_R $ (where $ E_R $ is the resonant energy
corresponding to $ \vt_G = \pi/4 $). Using (\ref{resca}) and (\ref{theta})
these conditions can be re-written as 
\begin{equation}
\nu^2 > {2 \delta \over \pi^2}  , \quad
f > \nu , \quad
f > \delta
\label{res.conditions}
\end{equation} 
where $ \delta = r_g \Delta m^2/ 4 E $ and $\nu = r_g \mu_t B $.

From (\ref{prob2}) we obtain (for the conditions at hand $ P_{LZ}
\simeq 0 $)
\begin{equation}
{ \nu \over \delta  + f } =  
\sqrt{ { P_t ( 1 - P_t ) \over \left| P_t - {1\over 2} \right|^2 }}
\label{mu.P}
\end{equation}
where $P_t = P( \nu_\mu \rightarrow \bar\nu_e )$ denotes the transition
probability and we have assumed $ \vt \simeq 0 $.

From (\ref{res.conditions}) and (\ref{mu.P}) it follows that 
\begin{equation}
f + \delta > f > \nu = ( f + \delta ) \sqrt{ { P_t ( 1 - P_t ) 
\over \left| P_t - {1\over 2} \right|^2 }}
\end{equation}
which required $ P_t > ( 2 + \sqrt{ 2 } ) / 4 $ or $ P_t < ( 2 - \sqrt{ 2 } )
/ 4 $; since these expressions are invariant under the replacement $ P_t
\rightarrow 1 - P_t $ we will assume  $ P_t > ( 2 + \sqrt{ 2 } ) / 4
\simeq 0.85 $.

For each fixed value of $P_t$ the relations above define a region in the
$ f - \delta - \nu $ space.
The quantity $f$ in (\ref{res.conditions}) takes, for the situations
under consideration, the values $ 10^{-4} - 1 $ depending on the production
region, with the larger values associated with regions closer to the
black hole horizon, (see Fig.~\ref{fig:f.2}). The allowed region in this
space for two values of $P_t$ are given in
Fig. \ref{fig:f.5}; the contour plot of $ \nu $ for various values of
$ \delta $ and $ P_t $ are presented in  Fig. \ref{fig:f.6}  for a
specific allowed value of $f$ 

\setbox4=\vbox to 170 pt{\epsfysize=4.5 truein\epsfbox[-130 -230 582 562]{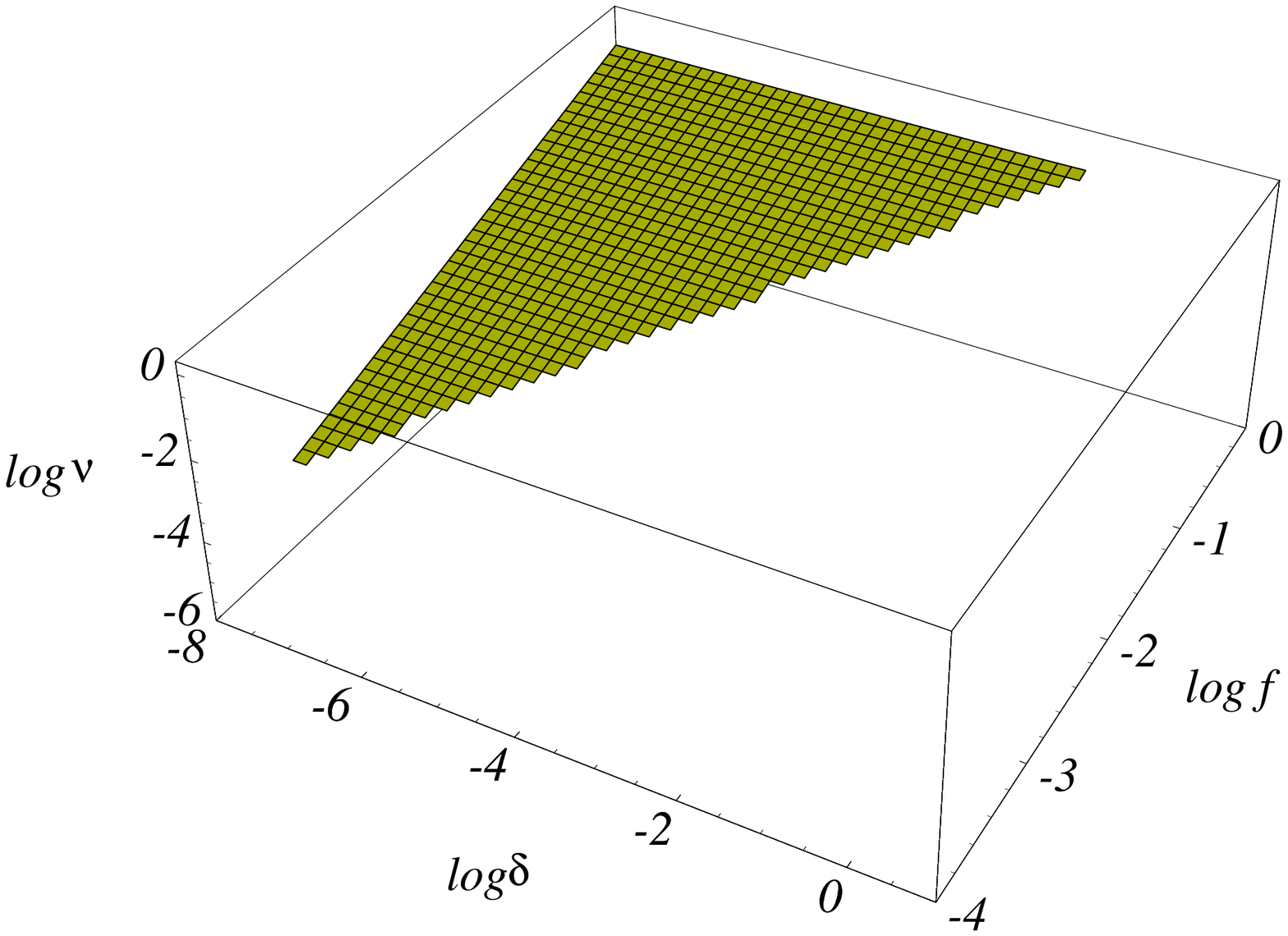}}
\setbox5=\vbox to 170 pt{\epsfysize=4.5 truein\epsfbox[5 -230 862 562]{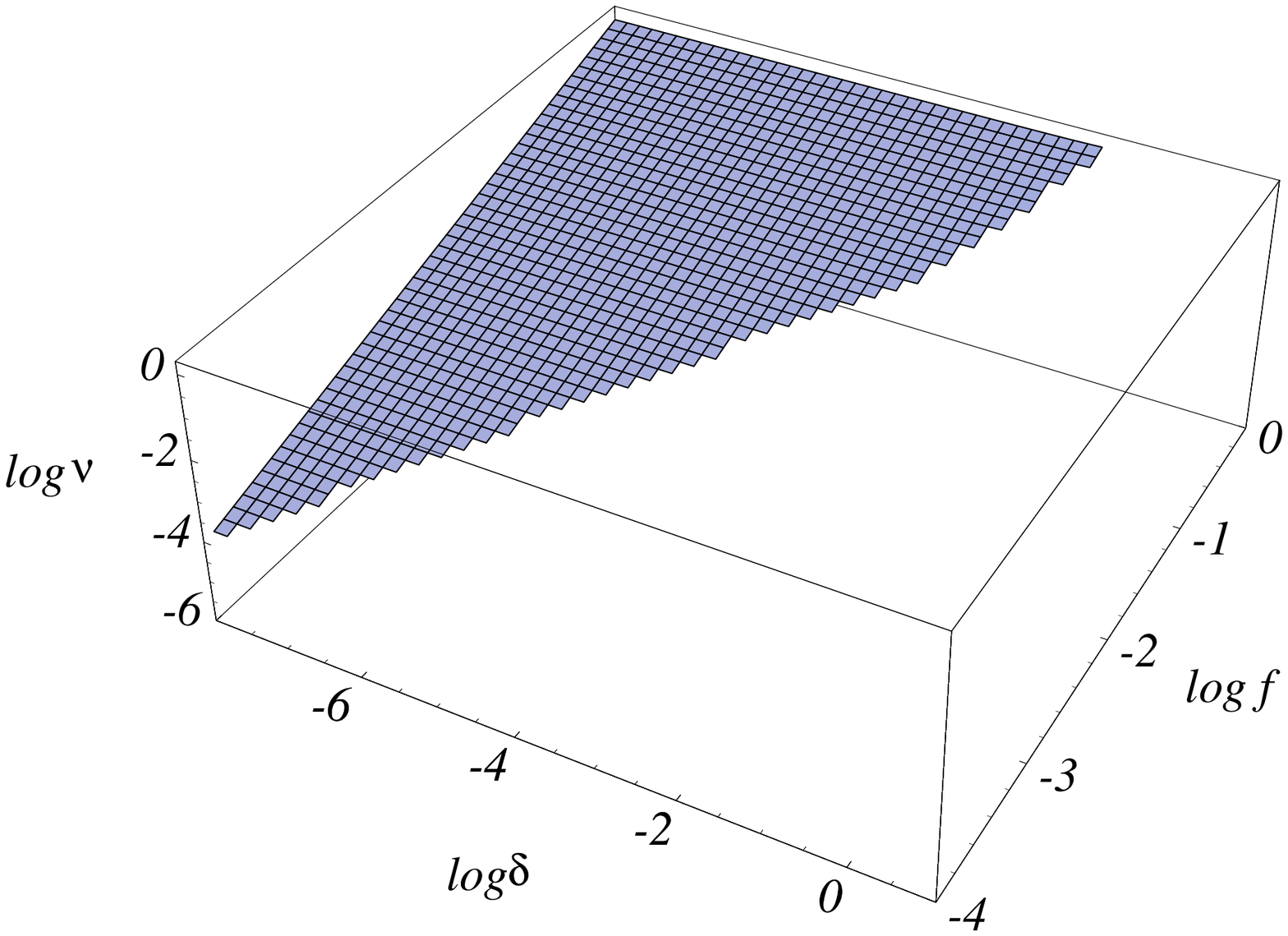}}
\begin{figure}
\centerline{\hfill \box4 \hfill \box5 \hfill}
\caption{Plot of $ \nu =  r_g \mu_t B $ 
as functions of
$f$ and $\delta = r_g \Delta m^2 /4 E$ for transition probabilities $P_t = 0.87$ 
(left) and $P_t = 0.99$ (right),  We used
$ M = 10^8 M_\odot$ and $ B = 10^4 $ G and imposed (\ref{res.conditions}).
\label{fig:f.5}}
\end{figure}

\setbox4=\vbox to 170 pt{\epsfysize=4.5 truein\epsfbox[-130 -230 582 562]{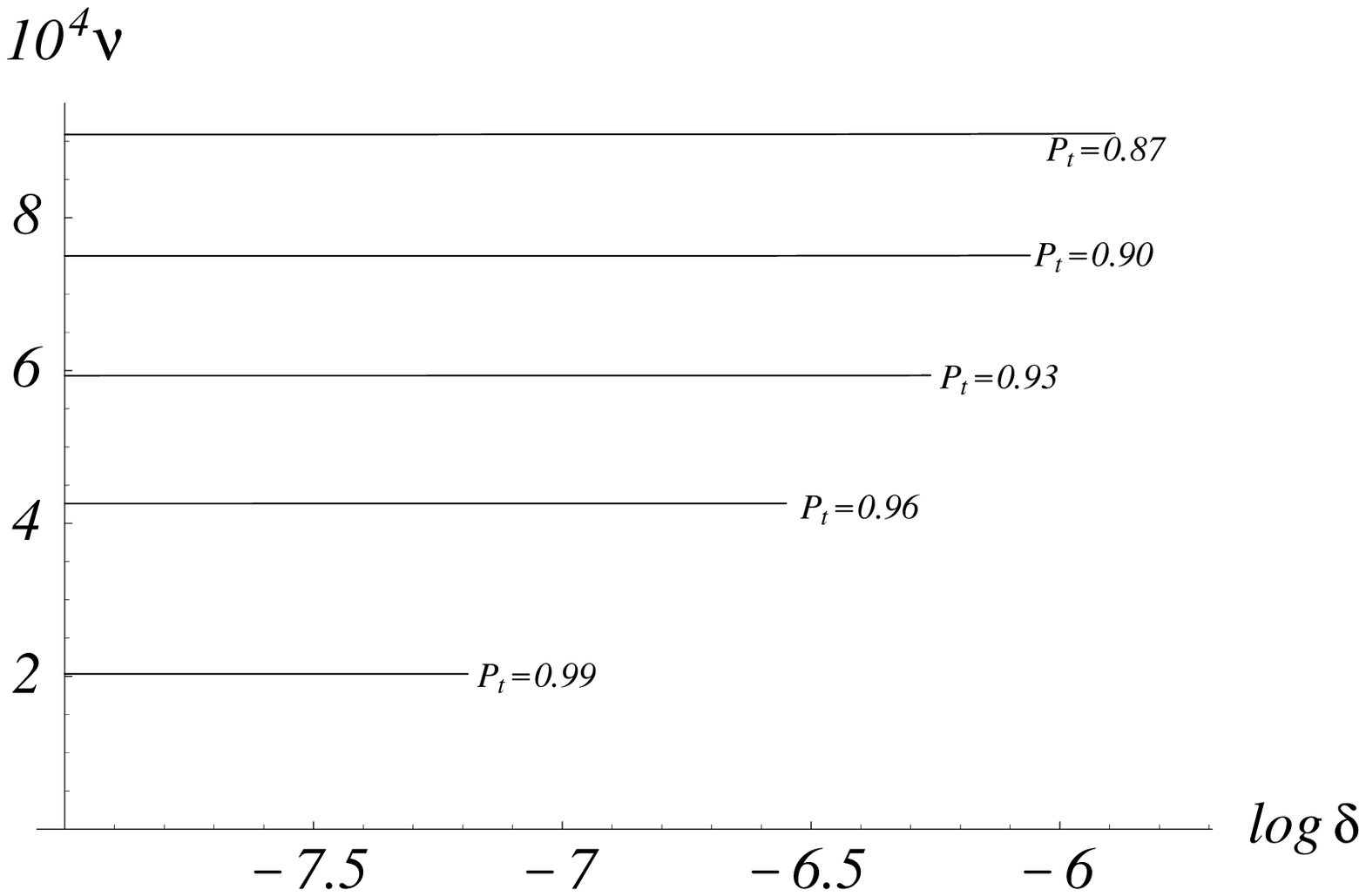}}
\begin{figure}
\centerline{ \box4 }
\caption{Plot of $ \nu =  r_g \mu_t B $
as  a function of $\delta = r_g \Delta m^2 /4 E$ for various values of the
transition probability $P_t $; we chose    
$ M = 10^8 M_\odot$, B $ = 10^4$ G, $ f = 10^{-3}$ and imposed (\ref{res.conditions}).
\label{fig:f.6}}
\end{figure}

As can be seen clearly from Fig.~\ref{fig:f.6}, for a reasonable value of
$f$ ($10^{-3}$), adiabatic resonant transitions occur for $ \mu_t $ in
the range $\sim 10^{-18}-10^{-17} \mu_B$.

The neutrinos will experience coherent
precession if the conditions $\nu > f $ and $ \nu >4 \delta $
are satisfied; in this case the transition and persistence probabilities
are $ \sim 1/2 $. For example, if $f = 10^{-3}$,
coherent precession will occur fod  $\mu_t > 10^{-17} \mu_B$ (for
the above values of $B$ and $ r_g$) assuming that $ \Delta m^2 < 4 E
\; \mu_t \; B $. For sufficiently large values of magnetic moment
coherent precession will be the dominant mechanism.

\section{Flux Modification due to Gravitational Oscillations.}

According to the spherical accretion model~\cite{sph} high energy
neutrinos are produced by proton acceleration at the shock.
Such acceleration is assumed to occur by 
the first order Fermi mechanism resulting in an
$ E^{-2} $ spectrum extending up to $ E^p_{max}$.
Estimates of expected neutrino fluxes from individual AGN are
normalized,
using this model, to their observed X-ray luminosities~\footnote{There
are varying results concerning the degree of 
proton confinement~\cite{equi}.
The various possibilities, however, alter only the low-energy neutrino
spectrum and does not affect our results.};
the position of the maximum neutrino energy, however, is uncertain because it
depends upon the parameters of the particular source.
According to the model we have used~\cite{1992}, the neutrino flux at
comparatively
 lower energies is strongly related to the X-ray flux; above 100 TeV
the flux depends strongly on the turnover in the primary photon
spectrum and one then needs to know the maximum proton energy $ E^p_{max}$.
The results of  calculations of Szabo and Protheroe~\cite{1992}
give an approximate formula for the muon neutrino 
$ (\nu_\mu + {\bar \nu}_\mu) $ flux in terms of $F_x$, the X-ray flux, 
and $E^p_{max}$
\begin{equation}
F_\nu \sim 0.25 F_x \exp(-20 E / E^p_{max}) E ^{-2}
\label{flux}
\end{equation}
where $F_x$ is a typical 2-10 Kev X-ray flux 
in $ {\rm erg}\; {\rm cm}^{-2} {\rm s}^{-1} $
and $ E$ is the neutrino energy; the flux is plotted in (Fig.~\ref{fig:f.7}). 
 
\setbox4=\vbox to 160 pt{\epsfysize=5.5 truein\epsfbox[0 -240 612 552]{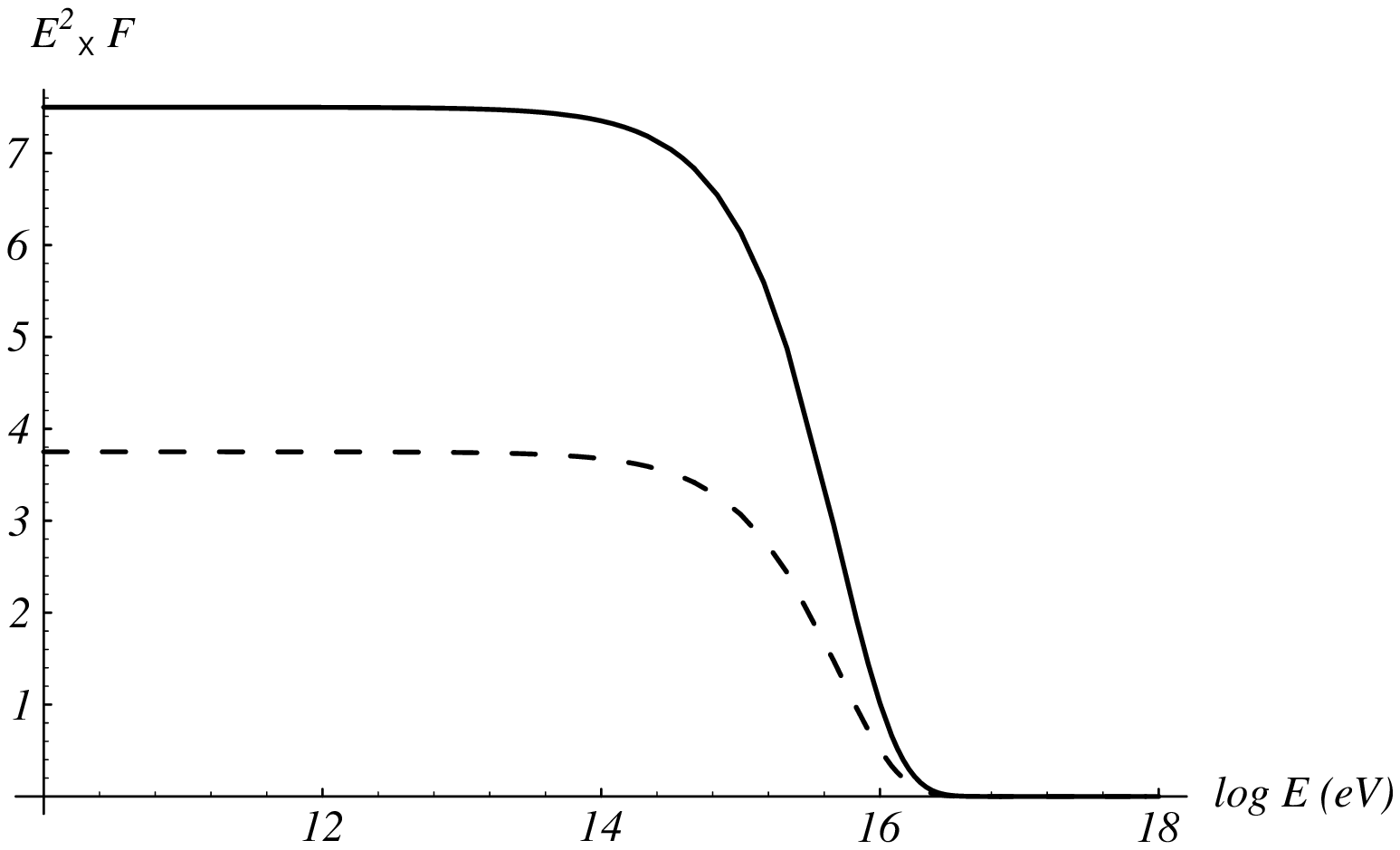}}
\begin{figure}
\centerline{\box4}
\caption{Plot of the $ (\nu_\mu + {\bar \nu}_\mu) $ (solid) and 
$ (\nu_e + {\bar \nu}_e) $ (dashed) fluxes, as given by (\ref{flux}),
as function of energy E when the X-ray flux equals
$F_x= 10^{-11} {\rm erg}  {\rm cm}^{-2} {\rm sec}^{-1} $. 
$ E^p_{max} = 10^{17} eV$ for AGN luminosity  $ 10^{45}$ ergs/sec, 
and $ \nu_\mu : \nu_e :: 2:1 $. \label{fig:f.7}}
\end{figure}

Taking  $F_x$ for a typical AGN to be 
$ 10^{-11} \,{\rm erg} {\rm cm}^{-2} {\rm sec}^{-1}$ 
(corresponding to for 3C273) and $E^p_{max} = 10^{17} eV$
for an AGN luminosity~\footnote{Assuming equipartition of
magnetic energy and radiation energy at shock,~\cite{equi} $E^p_{max} $is 
roughly proportional to $ \sqrt{L} $ . where $L$ denotes the AGN
luminosity.} $ L = 10^{45} {\rm ergs}/{\rm sec} 
$, the resulting  neutrino flux is plotted in figure 5.

As discussed in~\cite{raj} and~\cite{pakvasa}, the detectors 
AMANDA, Baikal and Nestor are sensitive to a wide range of
neutrino parameters and will be able to test a variety of
models of neutrino production in the AGN. For neutrino
energies above 1 TeV, measuring the ultra high energy muon flux permits an
estimation of the $ \nu_\mu + \bar{\nu}_\mu $ flux, for neutrino energies
above 3 PeV there is significant contribution to the muon rate due to
$ \nu_e $ interaction with electrons due to the W-resonance effect.
Also, as shown in~\cite{pakvasa}, the $\tau$ rate from double bang 
events can be used to measure the $ \nu_\tau + \bar{\nu}_\tau $ 
flux at energies 1 PeV and beyond.

We noted earlier that matter effects are negligible in the AGN
environment, but that gravity-induced resonances could cause
a modification of the neutrino flux of any given flavor. This effect
could cause an oscillation to $\tau $ neutrinos ~\footnote{ There is 
negligible $\tau $ neutrino production in the AGN environment according
to all standard models .} generating a significant flux of $ \tau $
neutrinos to which planned experiments will be sensitive in the
PeV range~\cite{pakvasa}. The observed $ \tau $ neutrinos would 
arise due to gravitationally induced resonance oscillations at the
AGN itself.

For the case of two flavors $ \alpha $ and $ \beta $
the observed neutrino flux for species $\alpha$,
$F_{\alpha}$, can be expressed in terms of the initially produced
neutrino fluxes $F_{\alpha}^0$ and $F_{\beta}^0$ as
\begin{equation}
F_{\alpha} =  [ 1 - P(\nu_\alpha \rightarrow \nu_\beta) ] F_{\alpha}^0
+ P(\nu_\beta \rightarrow \nu_\alpha) \, F_{\beta}^0 \label{fluxde}
\end{equation}
In order to calculate the fluxes in terms
of neutrino oscillations we have made
a number of simplifying but reasonable assumptions.

\begin{itemize}

\item The initial (production) fluxes are assumed to be in
the ratio $ \nu_\mu : \nu_e : \nu_\tau :: 2: 1: 0 $. If
charged and neutral pions are produced in equal proportions
simple counting leads to equal fluxes of $ \nu_\mu +
{\bar\nu}_\mu $ and photons. The flux of $ \nu_e +
{\bar\nu}_e $ equals half the flux of  $ \nu_\mu +
{\bar\nu}_\mu $. Ref.~\cite{raj} discusses the various
models AGN neutrino fluxes.

\item There are equal numbers of neutrinos and antineutrinos.

\item As discussed above and in ~\cite{mou} matter effects are
ignored. Resonances occur through the interaction of gravitational
and electromagnetic interactions (proportional to the transition
magnetic moment). We concentrate on the case where resonances are
present corresponding to small magnetic moments.

\item It is assumed that there is no large matter effects in the path from
AGN to the earth, so MSW effect does not cause oscillation for
neutrinos in transit either. Vacuum oscillations do occur for exceedingly
small $ \Delta m^2 $ (for a distance of $ \sim 100 $ Mpc and energy
1 TeV, $ \Delta m^2 < 10^{-19} {\rm eV}^2 $). Also the intergalactic
magnetic field $(\sim 10^{-6}{ \rm G })$ may cause spin transitions, but
this is again a small effects in the parameter range we consider.

\end{itemize}

Under these circumstances gravitationally
induced spin flavor oscillations are the only important process
which could cause neutrino transitions resulting in a modification of
the observed fluxes. Figures~\ref{fig:f.8}
show the decrease of $\mu$ neutrino flux  and the corresponding increase
in the $ \tau $ neutrino flux produced by 
resonant oscillations of $\mu$ neutrinos to $ \tau $ neutrinos; the effect increases with energy and
is very effective in the PeV range where $ \tau $ 
neutrinos can be detected relatively easily.

\setbox4=\vbox to 180 pt{\epsfysize=5.3 truein\epsfbox[-100 -240 512 552]{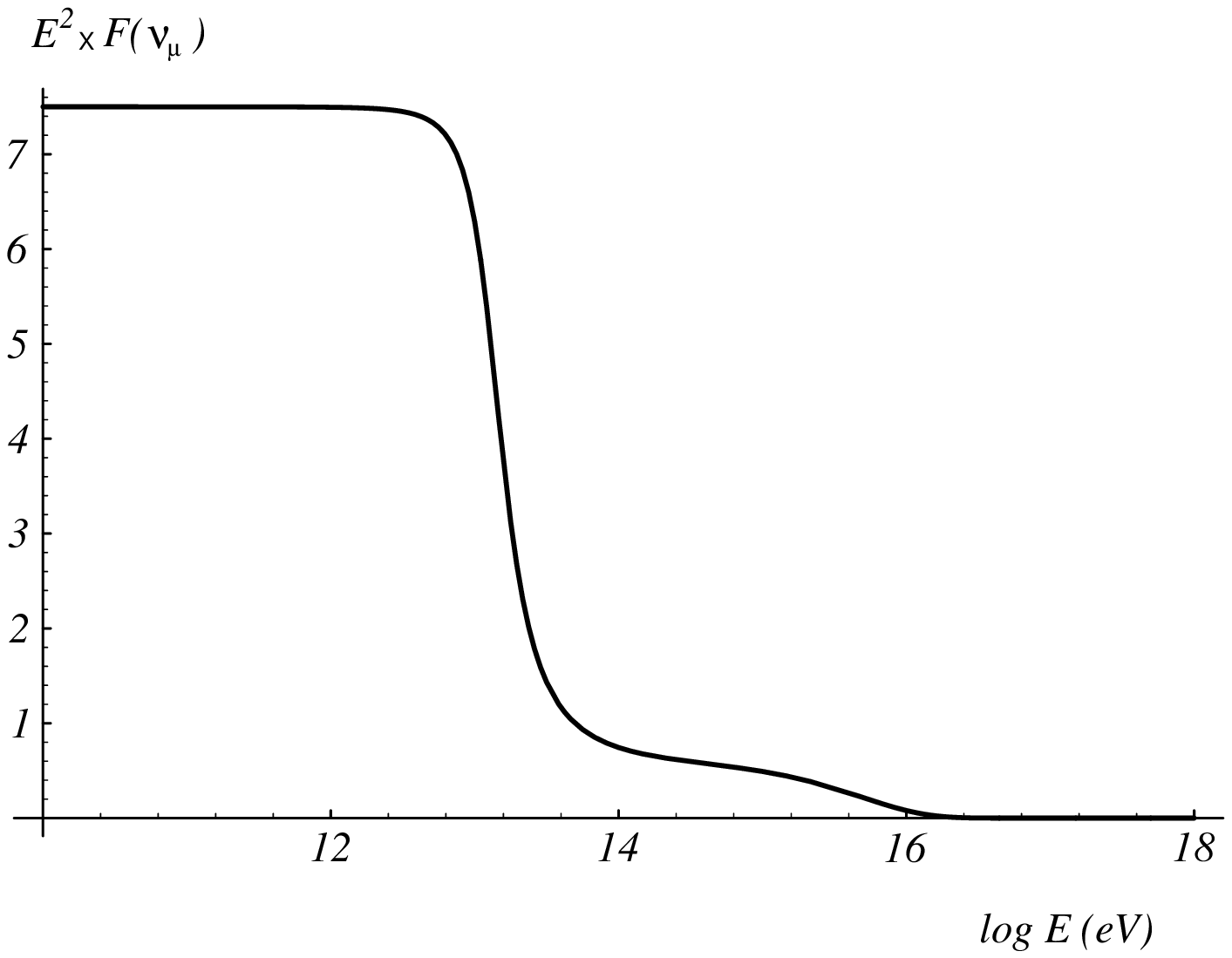}}
\setbox5=\vbox to 180 pt{\epsfysize=5.3 truein\epsfbox[0 -240 612 552]{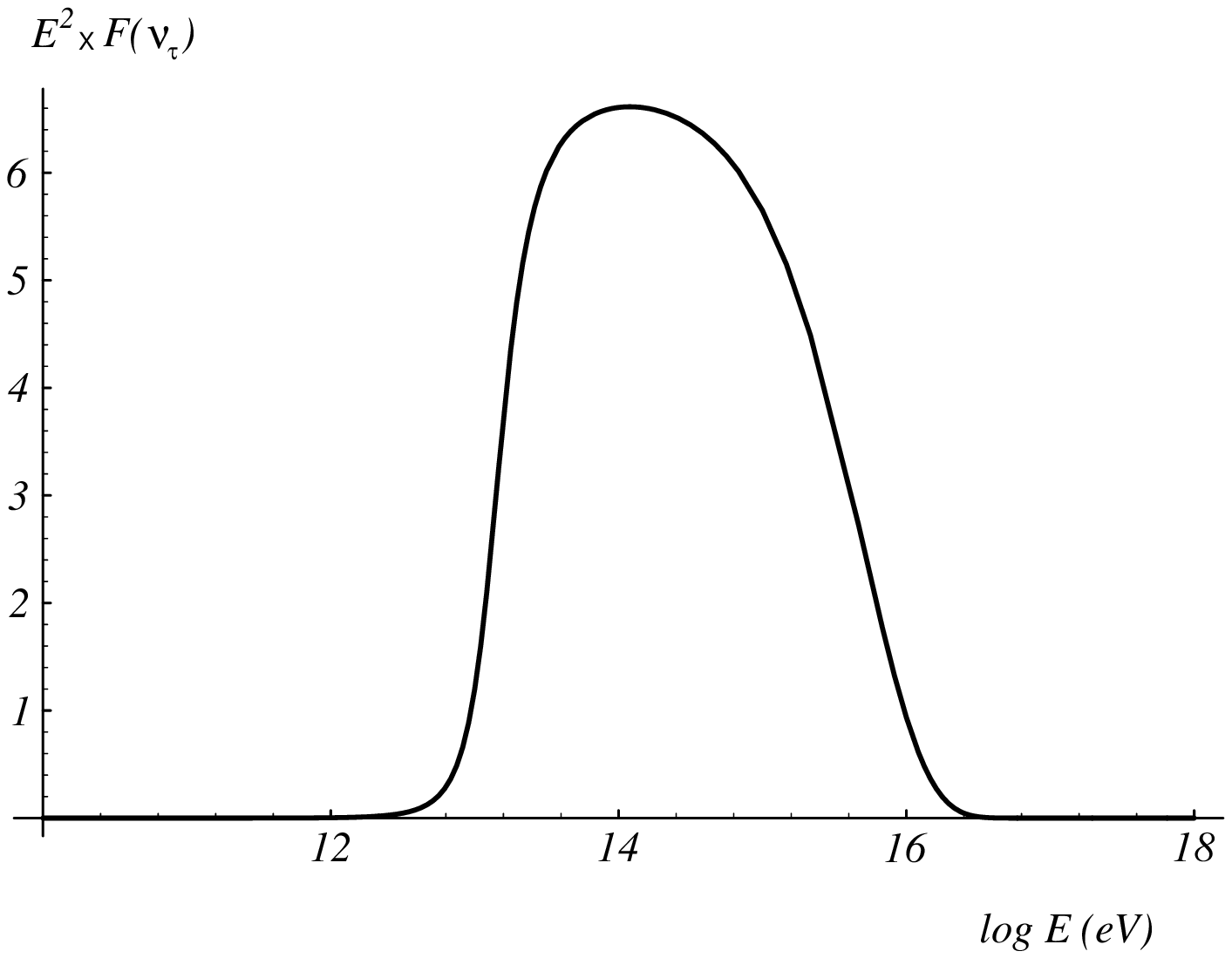}}
\begin{figure}
\centerline{\hfill \box4 \hfill \box5 \hfill }
\caption{Flux modification of $ \mu $ neutrino (left) and $ \tau $
neutrino fluxes due to
gravitational oscillations of the $ \nu_\mu - \nu_\tau $ system in AGN 
environment. $\Delta m^2 = 10^{-8} {\rm eV}^2$,
$\mu_t = 10^{-17} \mu_B $, $r/r_g = 6$,
$\theta=\pi/2$, $ \j=0.15 $, $\k=0.4$,
$ a/r_g=0.4$ and $ M = 10^8 M_\odot$. \label{fig:f.8}}
\end{figure}

The choice of $\mu_t = 10^{-17} \mu_B$ and $\Delta m^2 = 10^{-8} \rm eV^2$
for energy 1 PeV falls into the allowed region for gravitational
resonant oscillations (Fig.~\ref{fig:f.5}) 
satisfying conditions (\ref{res.conditions}).

\section{Conclusions}

We found that ultra-high energy Majorana neutrinos emanating from AGN are strongly affected by
gravitational and electromagnetic effects.
For typical values of gravitational current ($J_G$), which depends on
the allowed black hole and geodesic parameters, gravitational
resonant oscillations are found to occur for
energies in which future neutrino telescopes~\cite{pakvasa} will be 
sensitive provided $ \mu_t < 10^{-17}\mu_B $.
This therefore causes a significant increase of the corresponding $ \tau $ 
neutrino flux. Gravitational resonant oscillations
cause flavor/spin oscillation of such neutrinos where 
the transition neutrino magnetic moments
are small (Fig.~\ref{fig:f.5}).  For larger magnetic moments,
($ \mu_t > 10^{-17} \mu_B $) coherent precession will
dominate provided $\Delta m^2 < E \: \mu_t B $ \footnote{ The values of magneticmoment correspond to $ f = 10^{-3} $. 
For neutrinos produced in the immediate vicinity of the horizon $ f \sim 1 $ and resonances occur for $ \mu_t < 10^{-13} \mu_B $ while coherent precession dominates above this value.}. The increase in the $
\tau $ neutrino flux is, of course, accompanied by a corresponding
decrease in the $ \mu $ neutrino flux. The two cases discussed above can in principle be
differentiated by comparing the $ \tau $ and $ \mu $ neutrino fluxes
which would be equal for the case of coherent precession but markedly
different for the case of resonant oscillations (Fig.~\ref{fig:f.7} ).
However, transition efficiency is greatest for the minimum value of
the magnetic moment which corresponds to gravitational resonant oscillations.

We have restricted our calculations to the two neutrino flavor case. A
complete study should include at least three flavors and possibly four
(to consider the possibility of sterile neutrinos). In the case of solar
neutrinos the presence of more flavors can significantly alter the
predicted fluxes~\cite{panta}. However for the present situation where
the experimental information on the AGN neutrino flux is
quite limited, it is sufficient to determine the various effects and
their strengths by using a
a two flavor mixing description as we have done in our analysis.

\section{Acknowledgements}
 M.R. would like to thank Prof. Sandip Pakvasa for useful suggestions.
This work was supported in part by the US Department of Energy
under contract FDP-FG03-94ER40837.

\end{document}